\documentclass[prl,preprint,showpacs]{revtex4}
\usepackage{amsfonts,amsmath,amssymb,graphicx}
\usepackage{amsfonts}
\usepackage{amsmath}
\usepackage{amssymb}
\usepackage{graphicx}

\def\kbar{\protect\@kbar}
\def\@kbar{\relax \bgroup
\def\@tempa{\hbox{\raise.73\ht0
\hbox to0pt{\kern.25\wd0\vrule width.5\wd0 height.1pt
depth.1pt\hss}\box0}}\mathchoice{\setbox0\hbox{$\displaystyle
k$}\@tempa}{\setbox0\hbox{$\textstyle
k$}\@tempa}{\setbox0\hbox{$\scriptstyle
k$}\@tempa}{\setbox0\hbox{$\scriptscriptstyle k$}\@tempa}\egroup}
\input{tcilatex}

\begin{document}

\title{\textbf{Quantum Ratchets on Maximally Uniform States in Phase Space:
Semiclassical Full-Chaos Regime}}
\author{Itzhack Dana}
\affiliation{Minerva Center and Department of Physics, Bar-Ilan University, Ramat-Gan
52900, Israel}

\begin{abstract}
A generic kind of quantum chaotic ratchet is introduced, based on initial
states that are \emph{uniform} in phase space with the \emph{maximal possible%
} resolution of one Planck cell. Unlike a classical phase-space uniform
density, such a state usually carries a \emph{nonzero} ratchet current, even
in \emph{symmetric} systems. This quantum ratchet effect basically emerges
from the generic asymmetry of the state quasicoordinates in the Planck cell.
It is shown, on the basis of exact results, general arguments, and extensive
numerical evidence, that in a semiclassical full-chaos regime the variance
of the current over all the states is nearly proportional to the
chaotic-diffusion rate and to the square of the scaled Planck constant.
Experimental realizations are suggested.
\end{abstract}

\pacs{05.45.Mt, 05.45.Ac, 03.65.-w, 05.60.Gg}
\maketitle

\noindent 

Understanding quantum transport in generic Hamiltonian systems, which are
classically nonintegrable and exhibit chaos, is a problem of both
fundamental and practical importance. The study of simple model systems have
already led to the discovery of a variety of quantum-transport phenomena 
\cite{qc,d,er,hr,qr,qr1,qr2,qr3,qr4,qr5}, several of which have been
observed in atom-optics experiments \cite{er,qr4,qr5} and allow to control
and manipulate the quantum motion of cold atoms or Bose-Einstein condensates
in different ways.\newline

Recently, classical and quantum \emph{``ratchet''} transport in Hamiltonian
systems has started to attract a considerable interest, both theoretically 
\cite{hr,qr,qr1,qr2,qr3} and experimentally \cite{qr4,qr5}. Ratchets,
originally proposed as mechanisms for some kinds of biological motors and as
models for nanoscale devices \cite{rat}, are generally conceived as
spatially periodic systems with noise and dissipation in which an unbiased
(zero-mean) external force can lead to a directed current of particles due
to some spatial/temporal asymmetry. In classical Hamiltonian ratchets \cite%
{hr}, dissipation is absent and noise is replaced by deterministic chaos. A
basic general result \cite{hr} is that the average current of an initially 
\emph{uniform} density in phase space is always zero, even in the presence
of an asymmetry. This implies that the classical current of a chaotic region
may be nonzero only for an asymmetric system with a mixed phase space.
However, the corresponding quantized system can feature significant ratchet
effects also under \emph{full-chaos} conditions \cite{qr,qr1,qr2,qr3,qr4,qr5}%
. An important and apparently still open problem is the nature of these
effects in a semiclassical regime, in particular how precisely they vanish,
as expected, in the classical limit. A proper approach to this problem
should take into account the fact that the quantum current can be quite
sensitive to the \emph{initial state}, as indicated by recent exact results
in strong quantum regimes \cite{qr3}. Thus, the strength of the
quantum-ratchet effects should be globally measured by \emph{average}
characteristics of the current over a representative large set of initial
states. In order to understand these effects in a semiclassical regime and,
at the same time, to exhibit fundamental differences between classical and
quantum ratchets in the clearest way, it is natural to choose pure initial
states that are \emph{analogous as much as possible} to a classical
phase-space uniform density, for which ratchet effects are totally absent.
Initial states used until now to study quantum ratchets do not feature some
well-defined uniformity in \emph{all} phase-space directions. For example,
the often used pure momentum state \cite{qr,qr1,qr2} is uniform in position
but is infinitely localized in momentum.\newline

In this paper, we introduce a generic kind of quantum chaotic ratchet based
on initial states that are \emph{uniform} in phase space with the \emph{%
maximal possible} resolution of one Planck cell. Such a state is associated
with a phase-space lattice whose unit-cell area is $h$ and whose origin is
specified by ``quasicoordinates'' relative to a fixed Planck cell. We derive
a general exact expression for the quantum directed current carried by a
state as a function of its quasicoordinates. This current is usually \emph{%
nonzero}, also in \emph{completely symmetric} systems, due to the generic
asymmetry of the quasicoordinates in the Planck cell. Averages of the
current over quasicoordinates, corresponding to mixed states, are shown to
vanish. The quantum ratchet effect is globally measured by the \emph{variance%
} of the current over \emph{all} the maximally uniform states. We then
obtain from general arguments the following main result in a semiclassical
full-chaos regime: The variance of the current is nearly proportional to the
chaotic-diffusion coefficient and to the square of the scaled Planck
constant [see Eq. (\ref{Dh2}) below]. This implies, in particular, that a
system asymmetry is \emph{not} significant in this regime since it can
manifest itself only in the chaotic-diffusion rate. These results and
related issues, such as independence on the Planck-cell shape, are supported
by extensive numerical evidence. Finally, we briefly discuss possible
experimental realizations of the theory.\newline

We start by defining maximally uniform states in a phase space. Let $\hat{x}$
and $\hat{p}$ denote position and momentum operators, $[\hat{x},\hat{p}]
=i\hbar $. Then, $\hat{T}_{x}(a)=\exp (i\hat{p}a/\hbar )$\ and $\hat{T}
_{p}(b)=\exp (-i\hat{x}b/\hbar )$ are translation operators shifting $\hat{x}
$ and $\hat{p}$ by $a$ and $b$, respectively. A state $\left| \psi
\right\rangle $ is uniform on the phase-space lattice with unit cell formed
by $a$ and $b$ if it is invariant under application of $\hat{T}_{x}(a)$\ and 
$\hat{T}_{p}(b)$, up to constant phase factors: $\hat{T}_{x}(a)\left| \psi
\right\rangle =\exp (i\alpha )\left| \psi \right\rangle $ and $\hat{T}
_{p}(b)\left| \psi \right\rangle =\exp (-i\beta )\left| \psi \right\rangle $%
. This means that $\left| \psi \right\rangle $ is a simultaneous eigenstate
of $\hat{T}_{x}(a)$\ and $\hat{T}_{p}(b)$, implying that these operators
must commute. It is easy to show that $[\hat{T}_{x}(a),\hat{T}_{p}(b)]=0$
only if the unit-cell area $ab$ is a multiple of $h=2\pi \hbar $. Maximally
uniform states $\left| \psi \right\rangle $ correspond to the smallest unit
cell, i.e., the Planck cell with $ab=h$, and are given by \cite{kq} 
\begin{equation}
\langle x\left| \psi \right\rangle =\psi _{\mathbf{w}}(x)=\sum_{n}\exp (2\pi
inw_{2}/b)\delta (x-w_{1}-na),  \label{kq}
\end{equation}
where $\mathbf{w}=(w_{1},w_{2})$, with $0\leq w_{1}<a$ and $0\leq w_{2}<b$,
is related to the eigenphases $\alpha $ and $\beta $ above by $\alpha
=w_{2}a/\hbar $ and $\beta =w_{1}b/\hbar $.\ The quantities $w_{1}$ and $%
w_{2}$ are quasicoordinates specifying, respectively, the quasiposition and
quasimomentum of $\psi _{\mathbf{w}}(x)$ in the Planck cell: $w_{1}=x$ mod$%
(a)$, $w_{2}=p$ mod$(b)$. The states (\ref{kq}) form, for all $\mathbf{w}$,
a complete set and their $p$ representation is also a delta comb, at the
points $w_{2}+nb$ for all $n$ \cite{kq}. Thus, (\ref{kq}) is associated with
the phase-space lattice $(x,p)=\mathbf{w}+\mathbf{z}_{\mathbf{n}}$, where $%
\mathbf{z}_ {\mathbf{n}}=(n_{1}a,n_{2}b)$ for all $\mathbf{n}=(n_{1},n_{2})$%
. Pure momentum states may be viewed as the limit $a\rightarrow 0$ ($%
b=h/a\rightarrow \infty $) of the states (\ref{kq}).\newline

We consider the general quantum-ratchet systems described by Hamiltonians $%
\hat{H}(\hat{x},\hat{p},t)$ periodic in phase space $(\hat{x},\hat{p})$ and
in time $t$ \cite{hr}. Realistic such systems are known to exhibit robust
quantum-ratchet effects \cite{qr2}, see also below. Without loss of
generality, we assume that the period of $\hat{H}$ in both $(\hat{x},\hat{p}%
) $ is $2\pi $, implying a similar periodicity of the evolution operator $%
\hat{U}(\hat{x},\hat{p})$ in one time period. We shall derive a general
exact expression [Eq. (\ref{Ipe}) below] for the quantum-ratchet current
carried by a state (\ref{kq}) in terms of the quasienergy (QE) eigenvalues
and eigenstates of $\hat{U}(\hat{x},\hat{p})$. First, we present a
straightforward generalization of relevant results in Ref. \cite{d1}
concerning basic QE properties of operators $\hat{U}(\hat{x},\hat{p})$. A
generic value of the scaled Planck constant $\hbar /(2\pi )$ can be
approximated to arbitrary accuracy by a rational value, $\hbar /(2\pi )=q/N$
($q$ and $N$ are coprime integers). We then choose in (\ref{kq}) $a=2\pi
q_{1}/N_{1}$ and $b=2\pi q_{2}/N_{2}$ for some given integers $(q_{1},q_{2})$
and $(N_{1},N_{2})$ satisfying $q_{1}q_{2}=q$ and $N_{1}N_{2}=N$. Clearly, $%
\hat{U}$ commutes with both $\hat{T}_{x}^{N_{1}}(a)$ and $\hat{T}
_{p}^{N_{2}}(b)$, so that one can find simultaneous eigenstates of these
three commuting operators. It is easy to see that the general eigenstates of 
$\hat{T}_{x}^{N_{1}}(a)$ and $\hat{T}_{p}^{N_{2}}(b)$ can be expressed in
terms of (\ref{kq}) as follows: 
\begin{equation}
\Psi _{j,\mathbf{w}}(x)=\sum_{m_{1}=0}^{N_{2}-1}\sum_{m_{2}=0}^{N_{1}-1}\phi
_{j}(m_{1},m_{2};\mathbf{w})\psi _{w_{1}+m_{1}\hbar ,w_{2}+m_{2}\hbar }(x),
\label{qes}
\end{equation}
where the index $j$, $j=1,\dots ,N$, labels $N$ independent vectors of
coefficients, $\mathbf{V}_{j}(\mathbf{w})=\left\{ \phi _{j}(\mathbf{m};%
\mathbf{w}) \right\} $, for $\mathbf{m}\equiv (m_{1},m_{2})$ with $%
m_{1}=0,\dots ,N_{2}-1$ and $m_{2}=0,\dots ,N_{1}-1$. These coefficients are
determined by requiring (\ref{qes}) to be QE eigenstates of $\hat{U}$: $\hat{%
U}\Psi _{j,\mathbf{w}} (x)=\exp [-i\omega _{j}(\mathbf{w})]\Psi _{j,\mathbf{w%
}}(x)$, where $\omega _{j} (\mathbf{w})$ are $N$ QE bands. This leads to the
eigenvalue equation $\mathbf{\hat{ M}}(\mathbf{w})\mathbf{V}_{j}(\mathbf{w}%
)=\exp [-i\omega _{j}(\mathbf{w})]\mathbf{V}_{j} (\mathbf{w})$, where $%
\mathbf{\hat{M}}(\mathbf{w})$ is an $N\times N$ unitary matrix whose
elements can be expressed in terms of the Fourier coefficients of $\hat{U}(%
\hat{x},\hat{p})$, as in Ref. \cite{d1}.\newline

Now, the quantum-ratchet current in the $p$ direction for the initial state (%
\ref{kq}) is given by 
\begin{equation}
I(\mathbf{w})=\lim_{s\rightarrow \infty }\frac{1}{s}\int_{0}^{2\pi q_{1}}dx [%
\hat{U}^{s}\psi _{\mathbf{w}}(x)]^{\ast }\!\left( -i\hbar \frac{d}{dx}%
\right) \,[\hat{U}^{s}\psi _{\mathbf{w}}(x)]  \label{Ip}
\end{equation}
\noindent \hspace{0.03in}($s$ integer), where the integration range $2\pi
q_{1}$ is just the minimum common multiple of the periods $a=2\pi
q_{1}/N_{1} $ and $2\pi $ of (\ref{kq}) and $\hat{U}$, respectively; the
current in the $x$ direction is defined similarly. It is understood that one
has to replace $\psi _{\mathbf{w}}(x)$ in (\ref{Ip}) by a function $\tilde{%
\psi} _{\mathbf{w}}(x)$ which is normalized in $0\leq x<2\pi q_{1}$ and
which tends to $\psi _{\mathbf{w}}(x)$ in some limit. For example, 
\begin{equation}
\tilde{\psi}_{\mathbf{w}}(x)=\frac{\exp (iw_{2}x/\hbar )}{\sqrt{2\pi
q_{1}(2B+1) }}\sum_{n=-B}^{B}\exp [2\pi in(x-w_{1})/a]  \label{kqB}
\end{equation}
($B$ integer); essentially, $\tilde{\psi}_{\mathbf{w}}(x)\rightarrow \psi _{ 
\mathbf{w}}(x)$ as $B\rightarrow \infty $. Let us invert Eq. (\ref{qes})
using the completeness of the (orthonormal) eigenvectors $\mathbf{V}_{j}(%
\mathbf{w})$, i.e., $\sum_{j=1}^{N}\phi _{j}^{\ast }(\mathbf{m};\mathbf{w}%
)\phi _{j}(\mathbf{m} ^{\prime };\mathbf{w})=\delta _{\mathbf{m},\mathbf{m}%
^{\prime }}$: 
\begin{equation}
\psi _{\mathbf{w}}(x)=\sum_{j=1}^{N}\phi _{j}^{\ast }(\mathbf{0};\mathbf{w}%
)\Psi _{j, \mathbf{w}}(x).  \label{kqe}
\end{equation}
After inserting (\ref{kqe}) in (\ref{Ip}) and using the eigenvalue equation
above for $\mathbf{V}_{j}(\mathbf{w})$, we find that the dominant terms for
large $s$ are contributed by the functions $\partial \hat{U}^{s}(\hat{x},%
\hat{p} )/\partial \hat{x}\Psi _{j,\mathbf{w}}(x)$ given by $\sum_{\mathbf{m}%
}\bar{\phi} _{j,s}(\mathbf{m};\mathbf{w})\psi _{\mathbf{w}+\mathbf{m}\hbar
}(x)$, where the coefficients $\bar{\phi}_{j,s}(\mathbf{m};\mathbf{w})$ are
the components of the vectors $\partial \mathbf{\hat{M}}^{s}(\mathbf{w}%
)/\partial w_{1}\mathbf{V}_{j}(\mathbf{w })$ whose dominant behavior for $%
s\gg 1$ is $-is\partial \omega _{j}(\mathbf{w}) /\partial w_{1}\exp
[-is\omega _{j}(\mathbf{w})]\mathbf{V}_{j}(\mathbf{w})$. We then obtain the
exact result: 
\begin{equation}
I(\mathbf{w})=-\hbar \sum_{j=1}^{N}|\phi _{j}(\mathbf{0};\mathbf{w})|^{2}%
\frac{ \partial \omega _{j}(\mathbf{w})}{\partial w_{1}}.  \label{Ipe}
\end{equation}

Clearly, the current (\ref{Ipe}) is nonzero for generic values of $\mathbf{w}
$, even for a symmetric system, in which case $I(\mathbf{w})$ may generally
vanish only at symmetry points $\mathbf{w}$ of the bands $\omega _{j}(%
\mathbf{w})$. Let us assume, for definiteness, that the QE\ eigenvalues $%
\exp [-i\omega _{j} (\mathbf{w})]$ are nondegenerate at fixed $\mathbf{w}$.
This implies, using (\ref{qes}) and considerations similar to those in Ref. 
\cite{d1}, that: (a) Each QE band $\omega _{j}(\mathbf{w})$ is periodic in
both $w_{1}$ and $w_{2}$ with period $2\pi /N$. (b) $|\phi _{j}(\mathbf{0};%
\mathbf{w})|$ is periodic in $w_{1}$ and $w_{2}$ with periods $2\pi /N_{1}$
and $2\pi /N_{2}$, respectively; these periods define a unit cell $\mathcal{C%
}$ $q$ times smaller than the Planck cell. (c) $|\phi _{j}(\mathbf{m};%
\mathbf{w})|=|\phi _{j}(\mathbf{0};\mathbf{w}+ \mathbf{m}\hbar )|$. It
follows from (a), (b), and (\ref{Ipe}) that $I(\mathbf{w})$ is periodic in $%
\mathbf{w}$ with unit cell $\mathcal{C}$. Using (a), (c), (\ref{Ipe}), and
the normalization of $\mathbf{V}_{j}(\mathbf{w})$, $\sum_{\mathbf{m} }|\phi
_{j}(\mathbf{m};\mathbf{w})|^{2}=1$, we get that $\int_{0}^{2\pi
/N}dw_{1}\sum_{\mathbf{m}}I(\mathbf{w}+\mathbf{m}\hbar )=0$ at any fixed $%
w_{2}$. This result specifies general mixtures of the pure states (\ref{kq})
carrying a zero mean current and implies, in particular, that $\int_{%
\mathcal{C} }d\mathbf{w}I(\mathbf{w})=0$.\newline

The fluctuations of $I(\mathbf{w})$ around its zero mean are measured by the
variance: 
\begin{equation}
\left( \Delta I\right) ^{2}=\frac{q}{h}\int_{\mathcal{C}}d\mathbf{w}I^{2}(%
\mathbf{w}).  \label{var}
\end{equation}
The behavior of $\Delta I$ will now be studied in a semiclassical full-chaos
regime using arguments based on classical analogues of the states (\ref{kq}%
). For simplicity, we assume that $q=1$ ($\hbar =2\pi /N$), so that $%
\mathcal{C} $ is just the Planck cell, $N$ times smaller than the basic
torus of periodicity of the system, $T^{2}$: $0\leq x,p<2\pi $. The natural
classical analogue of the state (\ref{kq}) is the phase-space lattice $%
\mathbf{w}+\mathbf{z} _{\mathbf{n}}$ above, where now $\mathbf{z}_{\mathbf{n}%
}=(2\pi n_{1}/N_{1},2\pi n_{2}/N_{2})$ This can be restricted to a finite
lattice of $N$ points in $T^{2}$ with $n_{1}=0,\dots ,N_{1}-1$ and $%
n_{2}=0,\dots ,N_{2}-1$. Denoting by $p_{s}(\mathbf{w}+\mathbf{z}_{\mathbf{n}%
})$ the momentum evolving from initial condition $\mathbf{w}+\mathbf{z}_{%
\mathbf{n}}$ after $s$ time periods, the corresponding average classical
current is $I_{s}^{(\mathrm{c})}(\mathbf{w}+\mathbf{z} _{\mathbf{n}})=\Delta
p_{s}(\mathbf{w}+\mathbf{z}_{\mathbf{n}})/s$, where $\Delta p_{s} (\mathbf{w}%
+\mathbf{z}_{\mathbf{n}})=p_{s}(\mathbf{w}+\mathbf{z}_{\mathbf{n}%
})-w_{2}-n_{2}b$. The average current on the finite lattice, analogous to $I(%
\mathbf{w})$, is $\bar{I}_{s}^{(\mathrm{c})}(\mathbf{w})=\sum_{\mathbf{n}%
}I_{s}^{(\mathrm{c})}(\mathbf{w}+ \mathbf{z}_{\mathbf{n}})/N$. In analogy to 
$\int_{\mathcal{C}}d\mathbf{w}I(\mathbf{w})=0$ above, one has $\int_{%
\mathcal{C}}d\mathbf{w}\bar{I}_{s}^{(\mathrm{c})}(\mathbf{w})
/h=\int_{T^{2}}d\mathbf{z}I_{s}^{(\mathrm{c})}(\mathbf{z})/(4\pi ^{2})=0$,
where $\mathbf{z}\equiv (x,p)$ and the latter equality is the basic
uniformity result in Ref. \cite{hr}. We then obtain the classical analogue
of (\ref{var}) ($q=1$): 
\begin{equation}
\left( \Delta \bar{I}_{s}^{(\mathrm{c})}\right) ^{2}=\frac{1}{h}\int_{%
\mathcal{C}} d\mathbf{w}\left[ \bar{I}_{s}^{(\mathrm{c})}(\mathbf{w})\right]
^{2}=\frac{2}{Ns}\sum_ {\mathbf{n}}D_{s}(\mathbf{n}),  \label{varc}
\end{equation}
where $D_{s}(\mathbf{n})=\int_{T^{2}}d\mathbf{z}\Delta p_{s}(\mathbf{z}%
)\Delta p_{s} (\mathbf{z}+\mathbf{z}_{\mathbf{n}})/(8\pi ^{2}s)$ are
correlations of $\Delta p_{s}$ in phase space. For sufficiently large $s$, $%
D_{s}(\mathbf{0})$ ($\mathbf{z}_{\mathbf{0}}=\mathbf{0}$) is approximately
the chaotic-diffusion coefficient $D$, $\left\langle \left( \Delta
p_{s}\right) ^{2}\right\rangle _{T^{2}}\approx 2Ds $, while $D_{s}(\mathbf{n}%
)$ for $\mathbf{n}\neq \mathbf{0}$ ($\mathbf{z}_{\mathbf{n}}\neq \mathbf{0}$%
) should be negligible since it is expected to decay with $s$. Thus, (\ref%
{varc}) is nearly given by $2D/(Ns)$, showing how the full uniformity limit (%
$\Delta \bar{I}_{\infty }^{(\mathrm{c})}=0$) is approached by increasing the
lattice size $N$ and/or the length $s$ of a chaotic orbit which will then
visit almost uniformly the phase space. The quantum evolution should mimic
the classical one only up to a ``break time'' $s\sim s^{\ast }\sim 2\pi
/\Delta \omega =N$, where $\Delta \omega =2\pi /N$ is the mean spacing
between neighboring QE levels. Then, assuming that the limit in (\ref{Ip})
is essentially reached for $s\sim s^{\ast }\sim N\gg 1$ (semiclassical
regime), the variance (\ref{var}) will be approximately equal to $(\Delta 
\bar{I}_{s}^{(\mathrm{c})})^{2}\sim 2D/(Ns)$ with $s\sim N$: 
\begin{equation}
\left( \Delta I\right) ^{2}\sim \frac{2D}{N^{2}}=\frac{D\hbar ^{2}}{2\pi
^{2} }.  \label{Dh2}
\end{equation}

We have extensively checked the result (\ref{Dh2}) and related issues using
the generalized kicked Harper models \cite{d,d1} with $\hat{U}=\exp [-iL\cos
(\hat{p})/\hbar ]\exp [-iKV(\hat{x})/\hbar ]$, where $V(\hat{x})$ is a $2\pi 
$-periodic potential and $(K,L)$ are parameters. These are realistic models 
\cite{d,wg} (see below) and were shown recently \cite{qr2} to exhibit robust
quantum ratchet effects. The classical phase space is, in practice, fully
chaotic for sufficiently large $(K,L)$ and, as $K,L\rightarrow \infty $, $%
D/D_{\mathrm{ql}}\rightarrow 1$, where $D_{\mathrm{ql}}=\int_{0}^{2\pi
}dx[KdV(x)/dx]^{2}/2$ is the quasilinear value of $D$. We first verified
numerically for several potentials, parameter values, and Planck-cell shapes 
$(a=2\pi /N_{1},b=2\pi /N_{2})$, $N_{1}N_{2}=N$, that $D_{s}(\mathbf{n})$ in
(\ref{varc}) is indeed negligible for $\mathbf{n}\neq \mathbf{0}$ and $s=N$
if $D_{N}(\mathbf{0})$ is well converged to $D$; then, $(\Delta \bar{I}%
_{N}^{(\mathrm{c })})^{2}\approx 2D/N^{2}$ to a very good accuracy. The
quantum variance (\ref{var}) was calculated using the exact Eq. (\ref{Ipe}).
According to Rel. (\ref{Dh2}), $R\equiv (N\Delta I)^{2}/(2D_{\mathrm{ql}})$
should be approximately equal to $D/D_{\mathrm{ql}}$, independently of the
Planck-cell shape. This was supported by much numerical evidence. As
representative examples, Figs. 1 and 2 show $R$ versus $K$ in a completely
symmetric and strongly asymmetric case, respectively. In both cases, we
present results for two distinct factorizations of $N=121$: The ``most
uniform'' factorization $N_{1}=N_{2}=11$ (square Planck cell) and $N_{1}=121$%
, $N_{2}=1 $ (Planck cell elongated in the $p$ direction). Also shown is $%
D_{N}(\mathbf{0}) /D_{\mathrm{ql}}\approx D/D_{\mathrm{ql}}$. We see that
there is generally a reasonably good agreement between $R$ and $D_{N}(%
\mathbf{0})/D_{\mathrm{ql}}$, with only a weak dependence on the
factorization, especially in the symmetric case (Fig. 1). In this case,
discrepancies arise only around values of $K$ where small accelerator-mode
islands exist; these lead to the divergence of $D$ and to enhanced finite
values of $R$ and $D_{N}(\mathbf{0}) /D_{\mathrm{ql}}$. Fig. 3 shows plots
of $\Delta I$ versus $N$. The results agree very well with the $N^{-2}$ or $%
\hbar ^{2}$ behavior predicted by Rel. (\ref{Dh2}). The value of $\Delta I$
for sets of states less uniform than (\ref{kq}) was found to be
significantly larger than that in Figs. 1-3. For example, for the set of QE
states $\Psi _{j,\mathbf{w}}(x)$, with currents $I_{j}(\mathbf{w})=-\hbar
\partial \omega _{j}(\mathbf{w})/\partial w_{1}$, $\left
(\Delta I\right)
^{2}$ is nearly $N$ times larger than (\ref{Dh2}).\newline

In conclusion, we have introduced a generic kind of quantum ratchet in fully
chaotic systems, based on maximally uniform states in phase space and
described by average characteristics of the directed current over these
states. The fundamental source of this current is the generic asymmetry of
the quasicoordinates of a state in the Planck cell. A quantum ratchet effect
then arises also in completely symmetric systems from this single
unremovable source. According to Rel. (\ref{Dh2}), the average measure $%
\Delta I$ of the effect decays as $\hbar $ in a semiclassical regime (see
Fig. 3) and a system asymmetry can affect $\Delta I$ only in a
non-substantial manner, i.e., through the chaotic diffusion rate which does
not change significantly from completely symmetric to strongly asymmetric
systems (compare Figs. 1 and 2). We also found $\Delta I$ to be almost
independent on the Planck-cell shape, again in accordance with Rel. (\ref%
{Dh2}). The generalized kicked Harper models, used to illustrate our
results, are related to realistic systems, such as kicked charges in a
magnetic field \cite{d} and modified kicked rotors \cite{wg}, and may be
thus experimentally realizable by, e.g., atom-optics techniques. A state (%
\ref{kq}) can be well approximated by the finite superposition (\ref{kqB})
of plane waves (pure momentum states) whose number is small ($\sim 1$) if
the Planck cell is sufficiently elongated in the $p$ direction, as for the
second factorization in Figs. 1-3. A superposition of two plane waves was
used quite recently \cite{qr5} in experimental realizations of
quantum-resonance ratchets and superpositions of more plane waves may be
similarly prepared \cite{pc}. The new generic quantum ratchet effect
presented here should be then experimentally observable to some extent.%
\newline

This work was partially supported by the Israel Science Foundation (Grant
No. 118/05).

\newpage

\textbf{FIGURE CAPTIONS}\newline

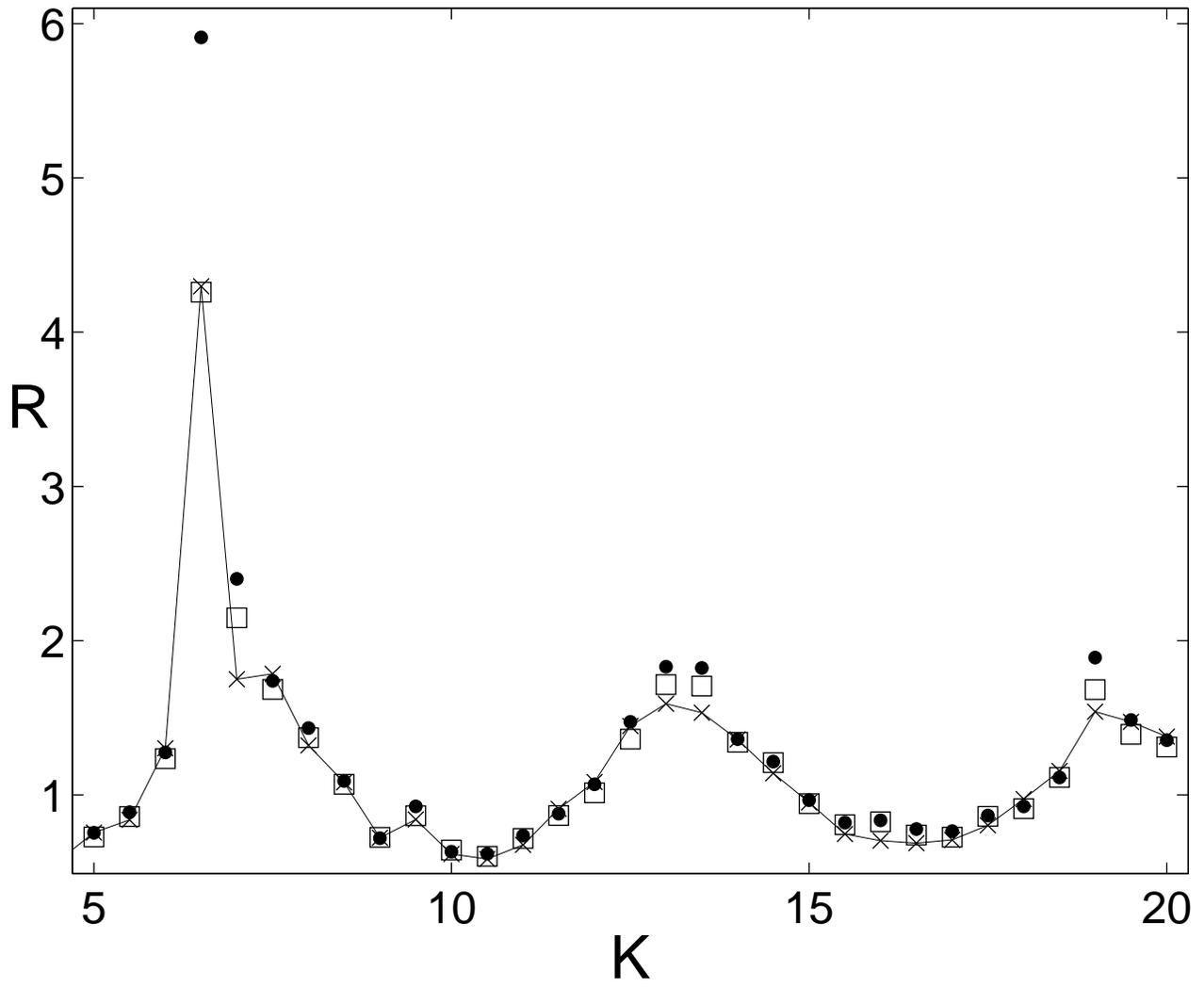
FIG. 1. Plots of $R\equiv (N\Delta I)^{2}/(2D_{\mathrm{ql}})$ versus $K$ in
the case of the symmetric kicked Harper model, with $V(x)=\cos (x)$ and $L=K$
($D_{\mathrm{ql}}=K^{2}/4$), for two factorizations of $N=2\pi /\hbar =121$: 
$N_{1}=N_{2}=11$ (squares) and $N_{1}=121$, $N_{2}=1$ (filled circles).
Crosses joined by line: The scaled diffusion rate $D_{N}(\mathbf{0})/D_{%
\mathrm{ql}}$ versus $K$ for this model ($N=121$). In all the figures, the
average over $\mathbf{w}$ in Eqs. (\ref{var}) and (\ref{varc}) was made on a
grid of $2500N$ points covering the Planck cell.\newline

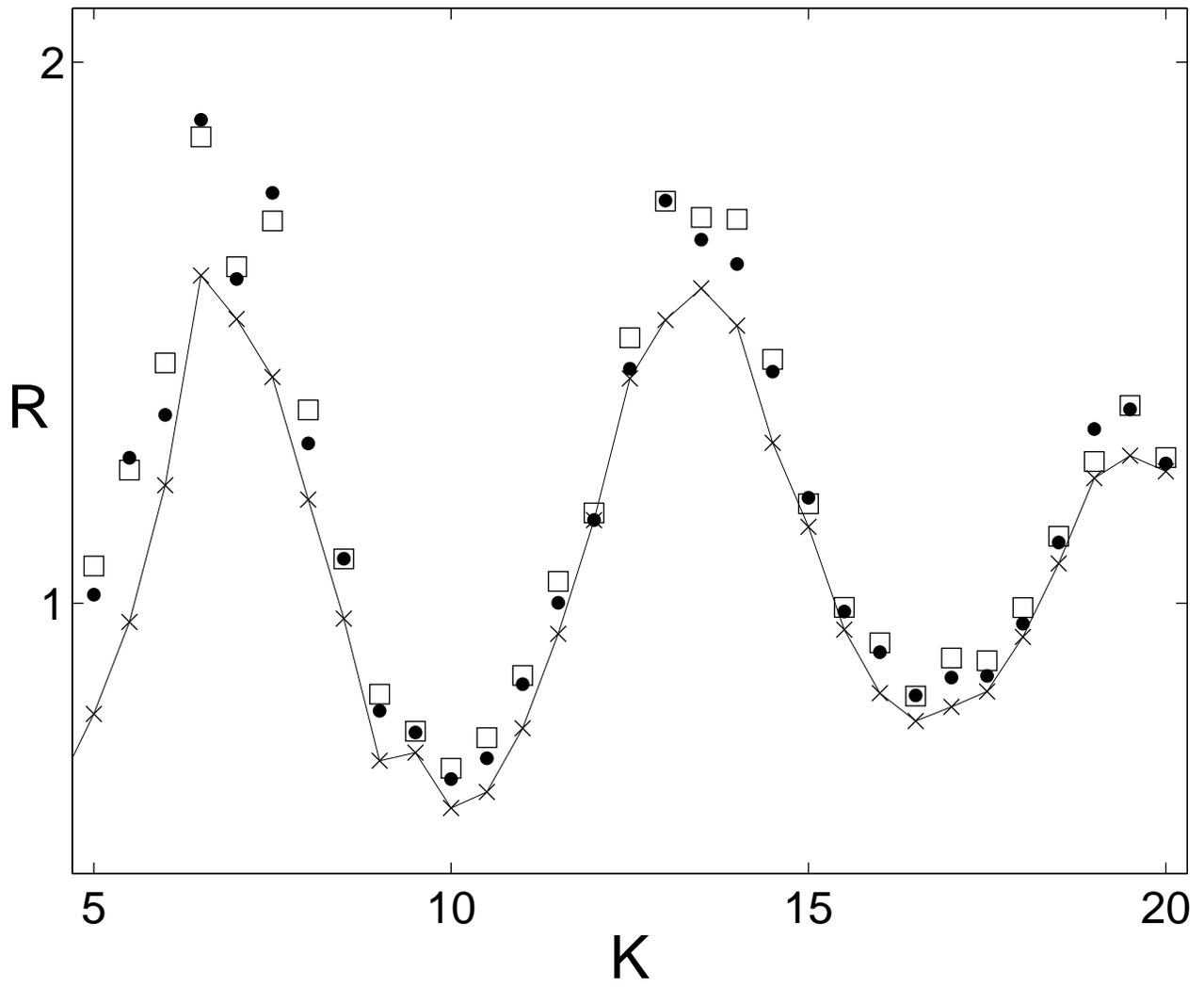
FIG. 2. Similar to Fig. 1 but for the strongly asymmetric kicked Harper
model \cite{qr2} with $V(x)=\cos (x)+\sin (2x)$ and $L=K/2$ ($D_{\mathrm{ql}%
}=5K^{2}/4$).\newline

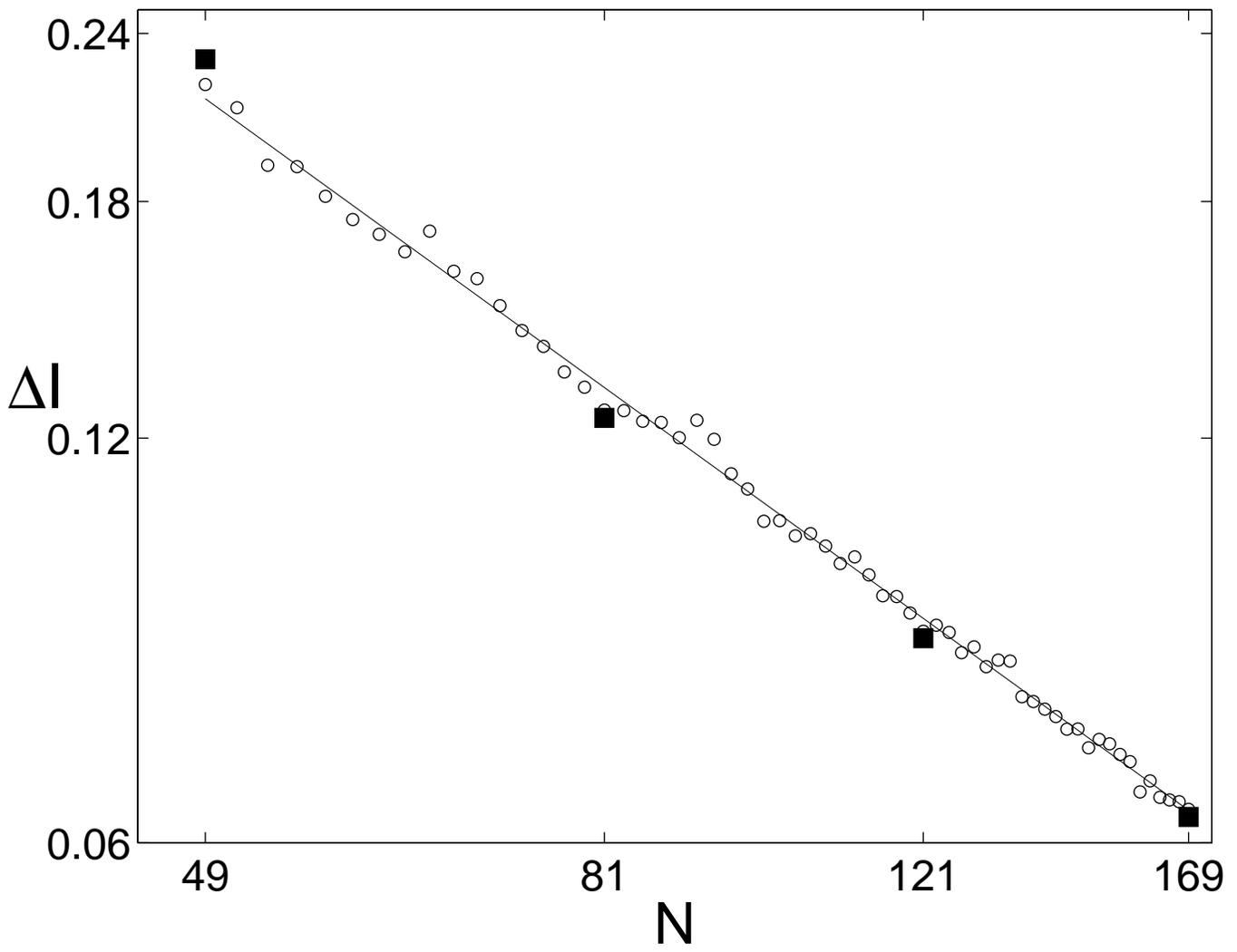
FIG. 3. Loglog plots of $\Delta I$ versus $N=2\pi /\hbar$ in the interval $%
49\leq N\leq 169$ ($N$ odd) for $N_1=N_2$ (filled squares) and $N_1=N$, $%
N_2=1$ (circles) in the case of the symmetric kicked Harper model with $%
V(x)=\cos (x)$ and $L=K=15$. The linear fit to the data has slope $-1.04\pm
0.06$ for $N_1=N_2$ and slope $-0.99\pm 0.01$ for $N_1=N$, $N_2=1$.\newline

\end{document}